\documentclass[aps,twocolumn,showpacs,preprintnumbers,amsmath,amssymb]{revtex4}

\usepackage{graphicx}
\usepackage{dcolumn}
\usepackage{bm}

\begin{document}


\title{Apparent phase transitions in finite one-dimensional sine-Gordon lattices}

\author{Sa\'ul Ares}%
\email{saul@math.uc3m.es}

\author{Jos\'e A.\ Cuesta}%
\email{cuesta@math.uc3m.es}

\author{Angel S\'anchez}%
\email{anxo@math.uc3m.es}

\affiliation{%
Grupo Interdisciplinar de Sistemas Complejos (GISC) and
Departamento de Matem\'aticas\\
Universidad Carlos III de Madrid, Avenida de la Universidad 30, 28911
Legan\'es, Madrid, Spain
}%
\homepage{http://gisc.uc3m.es}

\author{Ra\'ul Toral}
\email{raul@imedea.uib.es} \affiliation{ Institut Mediterrani
d'Estudis Avan\c cats (IMEDEA), CSIC-Universitat de les Illes
Balears\\ Campus UIB, 07071 Palma de Mallorca, Spain
}%
\homepage{http://www.imedea.uib.es}

\date{\today}

\begin{abstract}
We study the one-dimensional sine-Gordon model as a prototype of
roughening phenomena. In spite of the fact that it has been recently
proven that this model can not have any phase transition [J.\ A.\
Cuesta and A.\ S\'anchez, J.\ Phys.\ A {\bf 35}, 2373 (2002)],
Langevin as well as Monte Carlo simulations strongly suggest the
existence of a finite temperature separating a flat from a rough
phase. We explain this result by means of the transfer operator
formalism and show as a consequence that sine-Gordon
lattices of any practically achievable size
will exhibit this apparent phase transition
at unexpectedly large temperatures.
\end{abstract}

\maketitle

\section{Introduction}

More than fifty years ago, van Hove \cite{vanHove} proved that
true thermodynamic phase transitions, defined as singularities of
the free energy, could not occur in a class of one-dimensional
(1D) systems, a result later extended to lattice systems in the
same class by Ruelle \cite{ruelle}. In spite of the fact that the
conditions for van Hove's theorem to apply were clearly stated
\cite{vanHove} (see also \cite{lieb}), there is nowadays a very
general belief that 1D systems cannot exhibit phase transitions
unless they have long range interactions. This misinterpretation
of van Hove's mathematical results has been reinforced by the
abuse of Landau's \cite{landau} argument about the entropic
contribution of domain walls to the free energy. This argument,
being physically very intuitive and useful, is not a rigorous
result (assumptions such as a dilute concentration of domain walls
are made along the way) and, furthermore, it does not apply to
every 1D system. In fact, there are many examples of 1D systems
with true thermodynamic phase transitions
\cite{nagle,kittel,CW,burkhardt,dp} which, unfortunately, have
remained largely unnoticed.

In the more specific context of models of growth processes
\cite{BS,PV}, the unsustained belief
discussed above is often translated by saying that 1D interfaces are
always rough. This is actually not the case, as shown in the early
eighties with several examples \cite{CW,burkhardt}. Only recently,
two of us \cite{CS} have proven a theorem showing that a wide family of 1D models, including
the sine-Gordon model as a particular example, cannot have phase
transitions, this being, to our knowledge, the first time that a
rigorous proof of that kind has been given (some non-rigorous,
phenomenological
arguments in the same direction had been proposed earlier
\cite{FF}). However, as we will see below, this theorem turns out to be in conflict with some numerical simulation results which
seem to provide strong evidence supporting the existence of a roughening
phase transition in the 1D sine-Gordon model. In view of the fact that simulations are very often
the only way of studying a large class of models, it is
most important
to understand this contradiction in order to distinguish between
true and apparent phase transitions.

To the above end,
in this paper we focus on the 1D sine-Gordon model as a canonical example,
widely applicable and representative of the phenomenology of many
model systems \cite{schneider} (see also \cite{TS} for a review).
Thus, in Section \ref{sec:num} we give results of simulations
that suggest the existence of a phase transition at a
(not necessarily small) nonzero temperature. By means of a transfer
operator approach and using the probabilistic meaning of the corresponding
eigenfunctions, in Section \ref{sec:trans} we analyze the origin of
this behavior; from this analysis, we are able to conclude that
such apparent phase transitions will occur not only for lattice sizes
achievable within the present computational capabilities, but also
for very much larger lattices. Finally, in Section \ref{sec:disc} we discuss
the consequences of this result, which we believe are
relevant for computational studies where no analytical support exists. Furthermore, additional important implications of
our research for experimental studies of small systems far from
the thermodynamic limit are also considered.

\section{Numerical simulations}
\label{sec:num}

The 1D sine-Gordon model is defined by the following Hamiltonian:
\begin{equation}\label{hamil}
\mathcal{H}=\sum_{i=1}^N\big\{\frac{J}{2}(h_{i-1}-h_i)^2+
V_0[1-\cos(h_i)]\big\},
\end{equation}
where $N$ is the number of lattice nodes (or the system size), $J$
is the coupling constant and $-\infty<h_i<\infty$ is a real
variable on site $i$. We assume periodic boundary conditions $h_0\equiv h_N$. For visualization of our results, we
interpret $h_i$ as the height of a surface above site $i$ of the
lattice; then, the two terms of the Hamiltonian correspond, respectively, to
surface tension and to a local potential (of strength $V_0$) favoring
multiple values of 2$\pi$ for the height, 
representing that growth takes place preferentially by addition of
discrete units (layers). For surface growth
on two-dimensional (2D) substrate lattices, this 
interpretation has proven itself rather fruitful in the past (see
\cite{weeks,Falo,yo,sanbimo} and references therein). However, we
want to stress that the results we present in this paper are
independent of any specific interpretation one makes of $h_i$. In
fact, previous studies of the 1D sine-Gordon lattice \cite{schneider,GM}
were more interested in understanding the role of solitons in
statistical mechanics models \cite{TS} than in any particular
application.

In our study, we concentrate on two magnitudes in order to
characterize the model behavior: the surface width or roughness,
\begin{equation}
\label{rugosidad}
w^2=\left\langle{\frac{1}{N}\sum_{i=1}^N\lbrack{h_i-\bar{h}}\rbrack^2}\right\rangle,
\end{equation}
where
\begin{equation} \label{medh}
\bar{h}\equiv\frac{1}{N}\sum_{i=1}^Nh_i \end{equation} is the mean
height, and the height-difference correlation function,
\begin{equation}
\label{corre} C(r)=\left
\langle{\frac{1}{N}\sum_{j=1}^{N}\lbrack{h_{r+j}-h_j}\rbrack^2}
\right \rangle.
\end{equation}
Averages $\langle\cdots\rangle$ are to be understood with respect to a statistical weight given by the Gibbs factor, ${\rm e}^{-\mathcal{H}/T}$, at equilibrium at a temperature $T$.

The above defined are crucial quantities in the 2D version of the model. This exhibits a Kosterlitz-Thouless type phase transition from a low
temperature flat phase to a high temperature rough phase
\cite{Falo,sanbimo,CW2,knopf,nozi}. 
In the flat phase, small systems have a size dependent width whereas
the width of large systems is independent of the size. The crossover 
system size separating both regimes 
is closely related to the correlation length, which is 
finite in the low temperature phase,
and can be defined as the distance beyond which the
height-difference correlation function saturates. On the contrary,
in the rough phase the correlation length is infinite and, 
correspondingly, the
roughness increases (logarithmically in the 2D case) with the 
system size for all sizes, i.e., it is also infinite
in the thermodynamic limit. In 1D, the
theorem proved in \cite{CS} prohibits any phase transition, and at
all nonzero temperatures the system is in the rough phase, the
roughness increasing linearly with the system size.

In the lack of detailed analytical results, the statistical averages can be computed approximately by means of numerical simulations. This kind of analysis has become a routine tool for the study of the equilibrium properties of many models and a problem of interest is to extract from the numerical studies, necessarily performed in finite size lattices, the asymptotic behavior in the thermodynamic limit. We will show that a na{\"\i}ve extrapolation of the finite size results for the 1D sine-Gordon model can lead to erroneous results concerning the existence of a phase transition at a finite value of the temperature.

For our numerical study, and in order to assess the validity of our results, we have used two completely different procedures: Langevin dynamics and
parallel tempering Monte Carlo. The Langevin dynamics procedure has been widely used
in this context with very good results \cite{Falo,yo,sanbimo}, and
it consists in the numerical integration of the Langevin equation following from the Hamiltonian $\mathcal H$:
\begin{equation}
\frac{d h_i(t)}{d t}=-\frac{\partial \mathcal{H}}{\partial h_i(t)} +\eta_i(t),
\label{lange}
\end{equation}
where $\eta_i$ are Gaussian white noises
obeying the fluctuation-dissipation theorem at temperature $T$, i.e.,
\begin{equation}
\langle \eta_j(t') \eta_i(t) \rangle =2 T \delta_{ij}\delta(t-t').
\end{equation}
A major problem with the Langevin dynamics is the presence of systematic errors, in addition to the unavoidable statistical errors, due to the finiteness of the time step used in the numerical integration (in our studies, we have used a stochastic Heun method \cite{smt}). 

The second procedure, parallel tempering, is the one we have mostly
relied on. The reason is that it is a very efficient algorithm to
prevent the system from being trapped in local minimum energy
configurations. Parallel tempering requires any Monte Carlo method which generates representative configurations at a given temperature. In this case, we have implemented a heat bath algorithm \cite{toral}, in which new values $h'_i$ for the height at site $i$ are proposed according to the rule
\begin{equation}
\label{toral1}
h_{i}'=\frac{h_{i-1}+h_{i+1}}{2}+\xi \sqrt{\frac{T}{2J}},
\end{equation}
$\xi$ being a Gaussian random variable of zero mean and unit
variance, and are accepted with a probability $\min[1,{\rm e}^{-\delta\mathcal{H}/T}]$ with $\delta{\mathcal H}=-V_{0}[\cos(h_{i}')-\cos(h_{i})].$
The parallel tempering algorithm
considers then simultaneous copies of the system at different
temperatures, allowing exchange of configurations between them. The exchange occurs after enough configurations have been generated at each temperature for a time greater than the energy
autocorrelation time (see, e.g., \cite{newman,iba} for details).
This is particularly efficient for low
temperature configurations, which are most susceptible to being trapped
in metastable regions. All the results reported in this paper have been obtained by means
of this parallel tempering Monte Carlo algorithm, although we have checked that
Langevin dynamics produces the same results (quantitatively within
error bars).

Before proceeding with the discussion of the simulation results,
let us recall the theoretical background. As we have already
mentioned, the theorem proven in \cite{CS} implies that the system
is in the rough phase at all nonzero temperatures. This can be
interpreted in terms of general renormalization group arguments
(for a renormalization group study of the 2D sine-Gordon model, see
\cite{CW2,knopf,nozi}; see also \cite{BS,PV} for a more general
perspective): in the 1D model, fluctuations should be enough to effectively
suppress
the potential part of the Hamiltonian at any nonzero temperature,
leaving as the only relevant term the discrete gradient (surface
tension). In that case, the sine-Gordon model behaves effectively as the
Edwards-Wilkinson model \cite{EW}, whose associated Langevin equation (\ref{lange}) is simply the
discrete linear diffusion equation with additive noise, and all
the properties of interest can be calculated. This approach has
been very successful in characterizing the 2D sine-Gordon model behavior
\cite{sanbimo}, and particularly in locating the roughening
transition temperature. In our 1D case, it is easy to show that in
the Edwards-Wilkinson regime the roughness must scale linearly with the system
size for nonzero temperature \cite{ft}.

\begin{figure}
\vspace*{2mm}
\includegraphics[width=7.5cm]{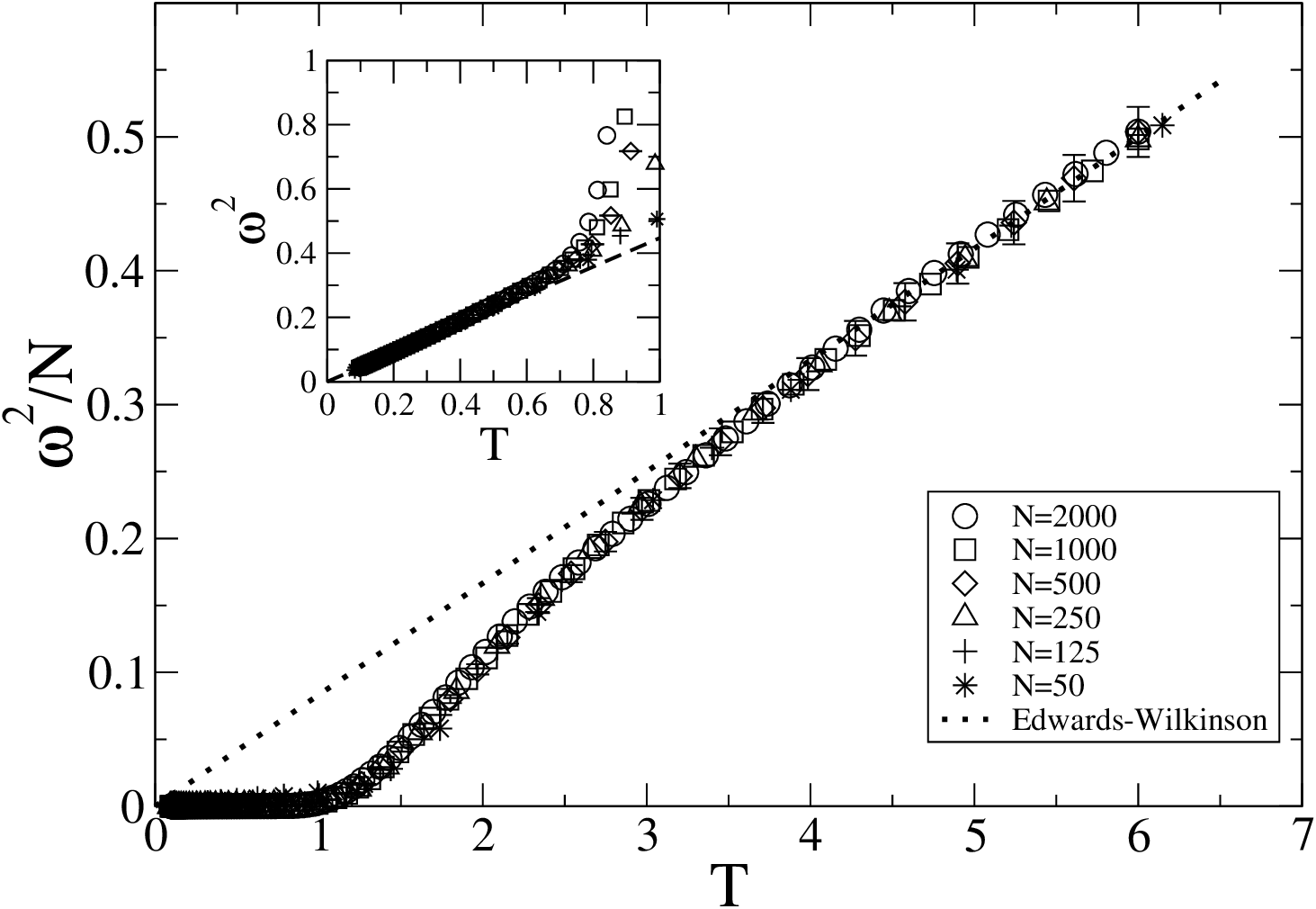}
\caption{\label{fig:rugosidad} Roughness scaled by the system size,
$\omega^2/N$,
vs temperature for the 1D sine-Gordon model. System sizes are as indicated in the
figure. The dotted line corresponds to the theoretical value $\omega^2/N=T/12$ obtained for
the Edwards-Wilkinson model. Inset: Zoom of the low temperature region
showing the lack of scaling. The dashed line corresponds to the parabolical
approximation discussed in the text, and is given by $\omega^2=T/\sqrt{V_0^2+
4V_0}$.}
\end{figure}
Figure \ref{fig:rugosidad} displays the simulation results for the
roughness for several system sizes as a function of temperature.
For the sake of definiteness we have chosen $J=V_0=1$ in Eq.\
(\ref{hamil}); other choices yield the same qualitative results.
Figure \ref{fig:rugosidad} shows that, as expected, the data tend asymptotically, for high values of the temperature, to the Edwards-Wilkinson result, $\omega^2/N=T/12$ \cite{ft}. This linear scaling of the roughness with system size indicates clearly that the surface is rough at high temperatures. 
However, the main plot in the figure indicates a clear change of behavior around a temperature $T\simeq 1$. In fact, as shown in the inset zoom,
at low temperatures ($T\lesssim 0.8$) the linear scaling dependence of the roughness with system size is lost and in that region the roughness becomes fairly 
independent of the system size, a behavior that according to the
preceding discussion would correspond to a flat phase.
To
obtain a theoretical prediction for low temperatures, we have
analyzed yet another linear model, in which the cosine term in the
Hamiltonian is substituted by a parabolic potential, $V_0(1-\cos(h_i))\to V_0h_i^2/2$. Such a model
is flat at all temperatures (basically because the parabolic
potential confines the surface to lie around its minimum) and, as
can be observed in Fig.\ \ref{fig:rugosidad}, agrees very well
with the numerical simulations at low temperatures.
We thus see
that, in spite of the fact that we know that no phase transition
can take place in this model, something very similar to a phase
transition from a flat to a rough phase appears in the simulations
for a temperature $T^*\simeq 0.8$.

\begin{figure}
\vspace*{6mm}
\includegraphics[width=7.5cm]{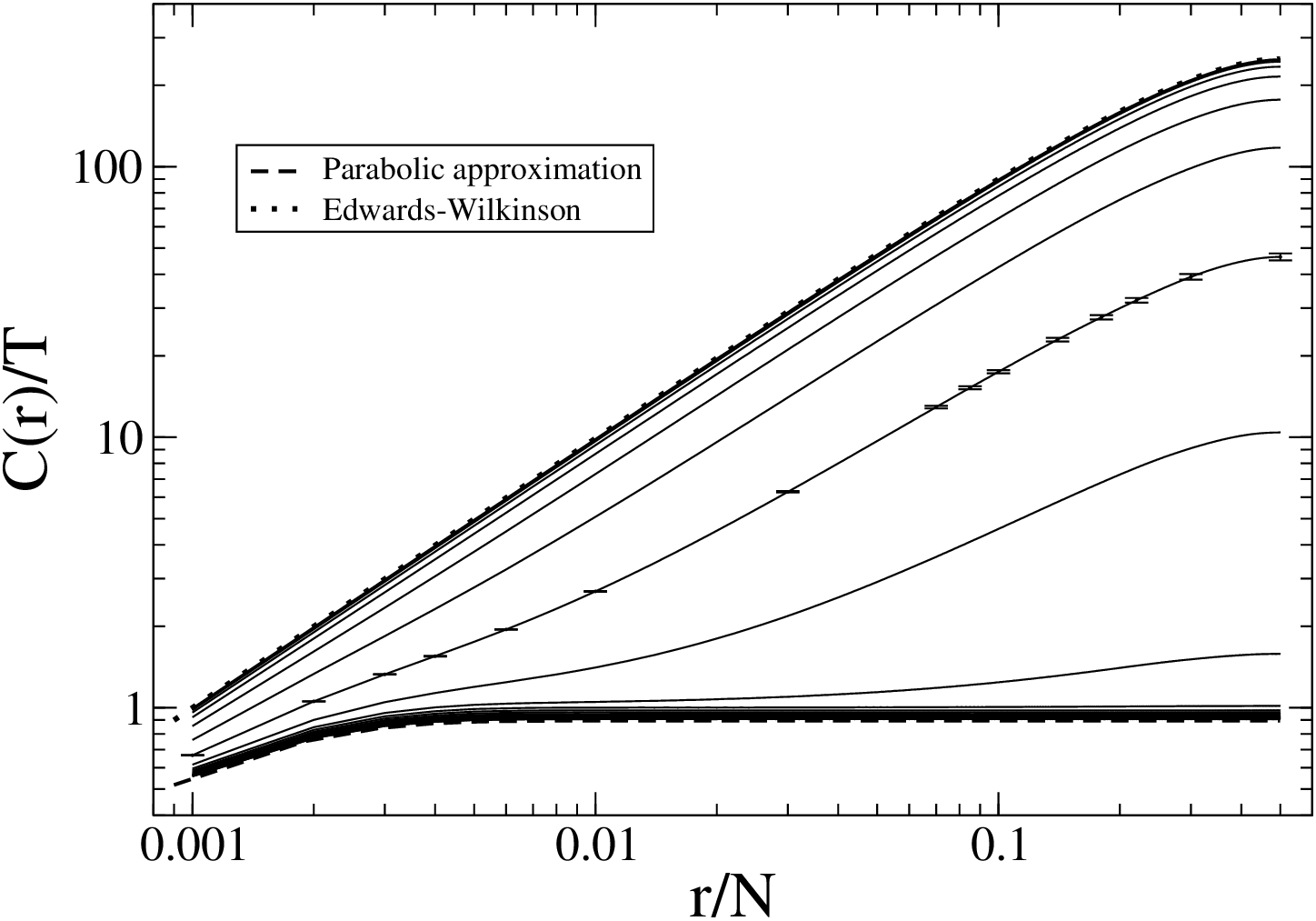}
\caption{\label{fig:correlacion} Log-log plot of the
height-difference correlation function scaled by
temperature vs distance scaled by the system size.
Bottom to top,
temperatures are 0.0956, 0.2407, 0.7029,
0.8115, 0.9896, 1.1689, 1.4819, 1.9016, 2.5562, 3.7044, 6. Also
plotted are the predictions of the Edwards-Wilkinson model and of a parabolic
approximation (see text). Error bars are typically as shown in one of
the curves.}
\end{figure}
The above indications are reinforced by looking at the correlation
function, shown in Fig.\ \ref{fig:correlacion}, where we can see
that for low temperatures there is a finite correlation length,
whereas for high temperatures the correlation extends as far as
half the system size (recall that we use periodic boundary
conditions). It can also be appreciated again that for high temperatures
the simulation results reproduce quite well the Edwards-Wilkinson prediction.
In the opposite limit, much as it occurs with the roughness, the 
correlation behaves like the parabolic approximation. 
We have also studied other
magnitudes, such as the specific heat, finding exactly the same
result as Schneider and Stoll \cite{schneider} that the specific
heat exhibits a maximum at a value somewhat higher than $T^*$. We
want to stress that all this is very reminiscent of the behavior
in 2D, where the maximum of the specific heat is interpreted as a
Schottky anomaly (see \cite{sanbimo} and references therein).
Hence, from the available
numerical evidence, we would be forced to conclude that there is a
roughening transition at a temperature $T^*$ in the 1D sine-Gordon model
were not for the theorem in \cite{CS}.

\section{Transfer operator approach}
\label{sec:trans}

In order to understand the numerical results, we will use the transfer
operator approach \cite{vanHove,dp,schneider,TS}; specifically, to
make the connection with the presentation in \cite{CS}, we rewrite
our Hamiltonian (\ref{hamil}) rescaling the $h_i$ variables by
a factor 2$\pi$, i.e.,
\begin{equation}\label{hamilr}
\mathcal{H}=\sum_{i=1}^N\big\{\frac{4\pi ^2 J}{2}(h_{i-1}-h_i)^2+
V_0[1-\cos(2\pi h_i)]\big\}.
\end{equation}
In this notation, the preferred values for $h_i$ are the integer
numbers, and the Hamiltonian is invariant under the transformation
$h_i\mapsto h_i+1$. This means that we can choose for convenience
$h_1\in\left(-\frac{1}{2},\frac{1}{2}\right]$ without loss of generality.
With this choice in mind, the corresponding partition function
becomes
\begin{equation}\label{funpar}
\mathcal{Z}_N(\beta)=\int_{-1/2}^{1/2}\textrm{d}h_1\int_{-\infty}^{\infty}\textrm{d}h_2\cdots\int_{-\infty}^{\infty}\textrm{d}h_N\textrm{e}^{-\beta \mathcal{H}},
\end{equation}
$\beta$ being the inverse temperature in units of the Boltzmann
constant. We now write $h_i=n_i+\phi_i$, with $n_i \in \mathbb{Z}$
and $-\frac{1}{2} < \phi_i \leq \frac{1}{2}$, with $i=1,\dots,N$.
Let us define
\begin{equation}\label{Vtheta}
V(\beta,\phi,\theta)\equiv\sum_{n=-\infty}^{\infty}\textrm{e}^{-\beta\frac{4\pi^2 J}{2}(n+\phi)^2}\textrm{e}^{-{\sf i}n\theta},
\end{equation}
a 2$\pi$-periodic function of $\theta$, and the operator
\begin{equation}\label{intop}
{\sf T}_{\beta,\theta}f(\phi)\equiv\int_{-1/2}^{1/2}\textrm{d}\phi'\mathcal{T}_{\beta,\theta}(\phi,\phi')f(\phi'),
\end{equation}
\begin{eqnarray}\label{tranop}
\mathcal{T}_{\beta,\theta}(\phi,\phi')&\equiv& V(\beta,\phi-\phi',\theta)
\times\\
&\times &
\exp\left\{-\frac{\beta}{2}V_0[2-\cos(2\pi\phi)-\cos(2\pi\phi')]\right\};
\nonumber\end{eqnarray}
with these definitions, the partition function can be written as
\begin{equation}\label{tranz}
\mathcal{Z}_N(\beta)=\frac{1}{2\pi}\int_{-\pi}^{\pi}\textrm{d}\theta\,\textrm{Tr}({\sf
T}_{\beta,\theta})^N.
\end{equation}
${\sf T}_{\beta,\theta}$ is called the transfer operator for this model.
Using the operator properties, it can be shown (see \cite{CS} for
details) that in the thermodynamic limit ($N\to\infty$)
\begin{eqnarray}\label{autos}
\textrm{Tr}({\sf T}_{\beta,\theta})^N&=&\sum_{n\geq 1}[\lambda_n(\beta,\theta)]^N\nonumber\\
&=&m(\beta,\theta)[\lambda_{\textrm{max}}(\beta,\theta)]^N[1+o(1)]
\end{eqnarray}
where $\lambda_n$ are the operator eigenvalues, necessarily real and isolated,
$\lambda_{\textrm{max}}$ is the maximum eigenvalue, necessarily positive,
and
$m(\beta,\theta)$ is its multiplicity, necessarily finite.
Finally, Laplace's method yields the free energy in the thermodynamic
limit:
\begin{equation}\label{freeen}
-\beta f(\beta)\equiv \lim_{N\to\infty}\frac{1}{N}\ln\mathcal{Z}_N(\beta)=\max_{-\pi\leq\theta\leq\pi}\ln[\lambda_{\textrm{max}}(\beta,\theta)].
\end{equation}

Based on this expression, in \cite{CS} it was proven that the
maximum of $\lambda_{\textrm{max}}(\beta,\theta)$ occurs at
$\theta=0$, and that $\lambda_{\textrm{max}}(\beta,0)$ is analytic
for $\beta>0$, which leads to the conclusion that the free energy
itself is analytic for all $\beta>0$ and, subsequently, that there
are no phase transitions in the model. However, that is not all
the information we can obtain from this approach, as we will now
show.

In \cite{schneider}, the squared modulus of the eigenfunction of the
largest eigenvalue is interpreted as the probability density for the
$h_i$ variables. As the transfer operator in \cite{schneider} is
different from the one we are using here, and for the sake of
completeness, we now proceed to show that we can resort to the
same interpretation here. Leaving out irrelevant constants, and
keeping in mind that as we have just said the only contribution to the free energy
comes from $\theta=0$, we can write
\begin{equation}
\mathcal{Z}_N(\beta)=\textrm{Tr}({\sf T}_{\beta,0})^N.
\end{equation}
We can now compute averages of functions
$g(\phi_j)$ in the following way \cite{schneider}:
\begin{equation}
\langle g(\phi_j) \rangle =\frac{\textrm{Tr}{\sf T_{\beta,0}}^jg(\phi_j){\sf T_{\beta,0}}^{N-j}}{\textrm{Tr}{\sf T_{\beta,0}}^N}.
\end{equation}
In the thermodynamic limit, it can be shown that 
the previous equation becomes 
\begin{equation}
\langle g(\phi_j) \rangle =\int_{-1/2}^{1/2}\textrm{d}\phi |\varphi_0(\phi)|^2g(\phi)
\end{equation}
where $\varphi_0(\phi)$ is the eigenfunction of the largest
eigenvalue. We thus see that, indeed, $|\varphi_0(\phi)|^2$ can be
understood as the probability density for the height to be in the
interval $\left(-\frac{1}{2},\frac{1}{2}\right]$ modulo $1$.

In order to apply this result, we must compute the eigenvalues and
eigenfunctions of the transfer operator.
This has to be done numerically: to this end, one has to discretize
the operator and transform
it into a matrix (see \cite{schneider,dp}, see \cite{thierry} for a detailed 
account). The advantage of the present formulation of the transfer
formalism is that integrals are to be carried out on a finite interval,
and therefore we do not need to introduce {\em ad hoc} any cutoff as in
the case of integrals on an infinite interval, thus eliminating one
possible source of error or inaccuracy. 
Specifically, in our numerical diagonalization procedure we have
used 2001$\times$2001 matrices; we have checked that increasing their
size does not change the results significantly. As another check, we
have computed the specific heat from the numerically computed eigenvalues,
finding perfect agreement with the output of the simulations.

\begin{figure}
\vspace*{2mm}
\includegraphics[width=7.5cm]{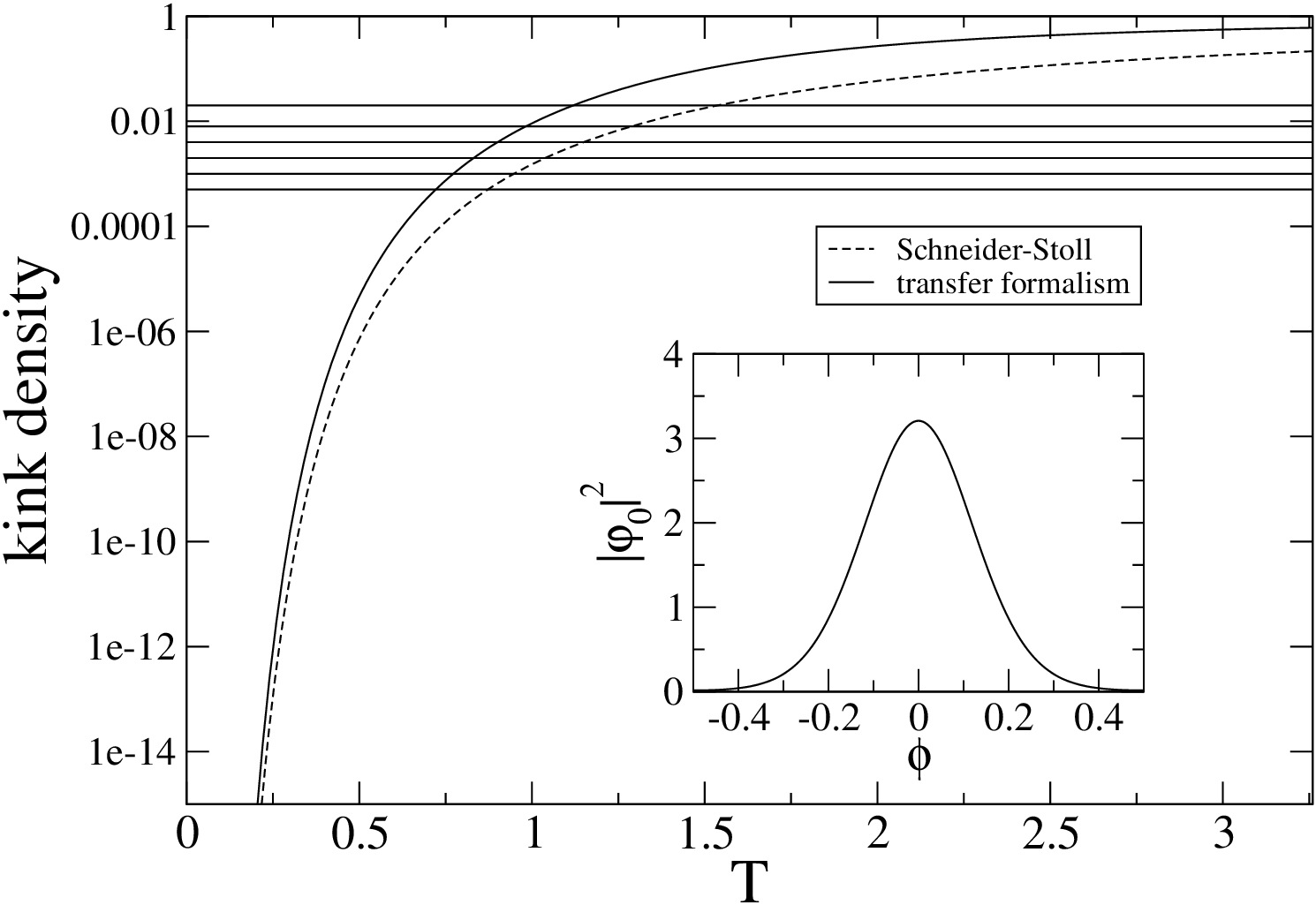}
\caption{\label{fig:everything} Kink density as estimated from the
probability that the value of the height goes over the potential
maxima at $\phi=\pm 1/2$ (see text). Horizontal lines correspond to the
inverses of the system sizes studied. For comparison, the approximate
result obtained by Schneider and Stoll \cite{schneider} is included
as a dotted line. Inset: squared modulus of the eigenfunction of the
largest eigenvalue at low temperature.}
\end{figure}
The inset in Fig.\ \ref{fig:everything} shows the squared modulus
of a typical eigenfunction at low temperature. The interpretation in
terms of probability density indicates that probable values for the
height lie close to the minimum of the potential, i.e., to $\phi=0$,
whereas values close to the maxima of the potential at $\phi=\pm 1/2$
are very unlikely to occur. We can associate the probability of
taking a value of $\phi=\pm 1/2$ with the probability of formation
of a kink or step, as once the height is at a maximum it can cross
over to the neighboring potential well, thus giving rise to a kink.
This interpretation suggests us to compare that estimate with the
inverse of the system sizes studied in this work, which is a
reasonable estimation of the probability of observing
a kink in our numerical system. Figure \ref{fig:everything} compares
both quantities, making clear that for every system size there is
a temperature at which the probability of formation of one kink
becomes smaller than the inverse of the system size. In fact,
the probability of formation of one kink decays extremely rapidly
below (orders of magnitude, note the logarithmic scale)
the crossing temperature.

\begin{figure}
\vspace*{2mm}
\includegraphics[width=7.5cm]{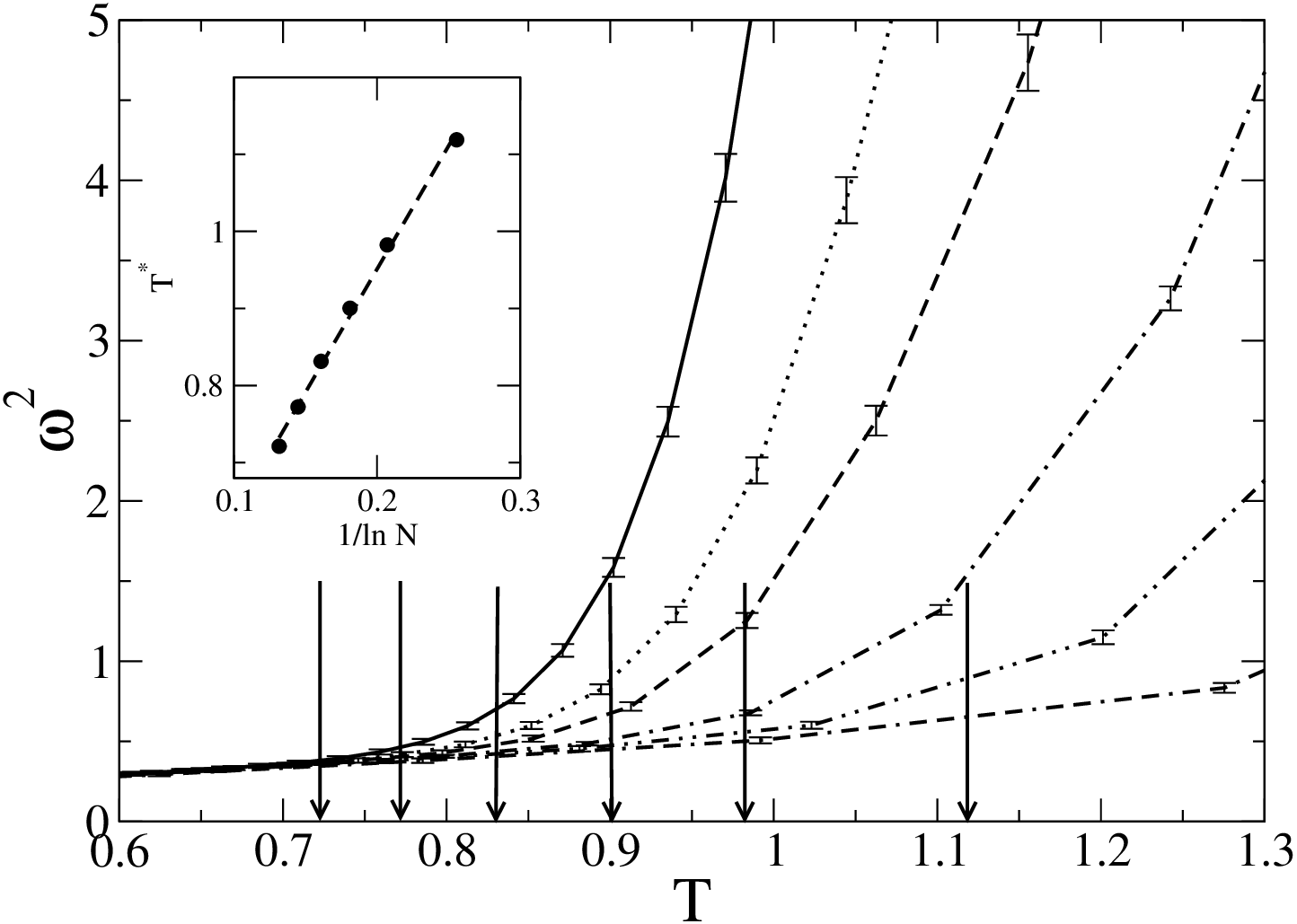}
\caption{\label{fig:compara} Roughness vs temperature for different
system sizes (right to left: $N=50, 125, 250, 500, 1000, 2000$).
The arrows mark the
temperature predicted by our criterion based on the eigenfunction
for each of those system sizes. Inset: estimates for $T^*$ vs $1/\ln N$.
Points are obtained from our criterion, line is a linear regression fit.}
\end{figure}
Following the discussion above, it is very natural then to
associate that crossing temperature with $T^*$, the temperature at which we
observe the apparent phase transition in our simulations. This is
very well confirmed by Fig.\ \ref{fig:compara}, in which the
roughness obtained from our numerical simulations is plotted along
with the temperatures predicted by our criterion above. We note in
passing that this result allows us to understand the reason why
the crude parabolic approximation coincides so well with the
numerical results at low temperatures: the hypothesis underlying
the approximation is that kinks do not form and the whole surface
lies close to a single potential minimum. Indeed, we have
seen that the probability of kink formation at low temperatures
is certainly negligible. 

Summarizing, we have support for the criterion, as confirmed by the comparison
of the transfer eigenfunction approach to the results of the numerical
simulations, that the apparent transition temperature $T^*$ coincides with the
one that yields a kink formation probability smaller than the inverse of the
system size.
This criterion allows us to make the following quantitative prediction: for a system of size $N$, there is an
apparent flat-rough phase transition at a temperature $T^*\sim
1/\ln N$, as shown in the inset of Fig.\ \ref{fig:compara}. This
can be easily understood if we realize that as kink formation is
an activated phenomenon, the kink density follows an Arrhenius
type law (with corrections, see \cite{schneider}). The dependence
of $T^*$ on $N$ follows from our criterion by imposing the proportionality of
the kink density and $1/N$. We
can now estimate the size of the system needed in order to observe
the rough phase all the way down to any given temperature, $T^*$.
Taking as an example $T^*=0.1$, which is certainly not small, we
find that lattices of the order of $10^{30}$ sites are required to
ensure a reasonable chance that kinks are formed during the
simulation and the rough phase is observed for $T>T^*$; we would
still find such an exceedingly large system in the flat phase for
$T<T^*$. We have thus shown clearly that systems of any
practically achievable size will always exhibit an apparent phase
transition at a temperature far from $T=0$.

\section{Conclusions}
\label{sec:disc}

In this paper, we have studied analytically and numerically, by
Langevin dynamics and mostly by parallel tempering Monte Carlo
simulations, the 1D sine-Gordon model. We have found in the simulations
that there exists a temperature at which an apparent
roughening phase transition takes place. We have shown that it
is possible to understand the contradiction of such phenomenon
with the theorem that prohibits phase transitions in this model
\cite{CS} through the analysis of the eigenfunctions of the
corresponding transfer operator. The interpretation of these
functions as probability densities makes clear that lattices
of any finite size will always show an apparent phase transition,
because the probability that kinks are formed becomes negligible
below certain temperature. We have also seen that even in
extremely large lattices the apparent transition occurs at
temperatures far from zero and, in fact, $T^*\sim 1/\ln N$.

The results summarized above are relevant in a much broader
context, basically in two directions. First, our conclusion should
be kept in mind when analyzing the outcome of numerical
simulations of models about which there is little or none
analytical information. Were not for the fact that we know that
such a phase transition is not possible, we would have concluded
from our simulations that the 1D sine-Gordon model presents a roughening
phase transition. In fact, simulations for the 2D sine-Gordon model yield
results very similar to those presented here \cite{sanbimo}, although in that
case we have a true phase transition according to several
approximate calculations including renormalization group results
\cite{weeks,CW2,knopf,nozi}.
It is important to realize that finite size analysis, which in
principle could signal that the transition goes to $T=0$ with
system size, becomes questionable if the values for the apparent
transition at the sizes amenable within computational capabilities
are still very far from $T=0$. In addition, in our model,
approximate analytical results in the low temperature limit
support the existence of the non-existent phase transition.
Therefore, we conclude that one has to be extremely careful with
claims of this kind.

The second direction that our work points to is related to the
very nature of phase transitions. True thermodynamic phase
transitions, understood as singularities of the free energy or its
derivatives, can only take place in the thermodynamic limit, and
no such transitions occur in finite size systems. Hence, the
apparent phase transition we see in our simulation can indeed be
thought of as a true transition in the context of finite systems.
What is more important, similar phenomenology is bound to arise in
small, mesoscopic systems, certainly far from any thermodynamic
limit one can think of. As systems of that scale become more and
more relevant both for theoretical and for applied reasons, the
question of the definition and nature of phase transitions gains
importance. In this respect, this work hints that
non-thermodynamic transitions may well be physically existent, or,
alternatively, that computations and results in the thermodynamic
limit do not represent well the fate of large but finite systems,
even of very large, mesoscopic ones.

\begin{acknowledgments}
S.A.\ and A.S. want to thank IMEDEA and Universitat de les Illes
Balears for their hospitality during the progress of this work.
S.A.\ and R.T.\ thank Jo\~ao M.\ V.\ P.\ Lopes for hospitality and
discussions at Centro de F\'\i sica do Porto. This work has been
supported by the Ministerio de Ciencia y Tecnolog\'\i a of Spain and FEDER 
through grants BFM2000-0004 (J.A.C.), BFM2000-0006 (S.A. and
A.S.), BFM2001-0341-C02-01 and BFM2000-1108 (R.T.).
\end{acknowledgments}

\end{document}